\documentstyle[aps,epsfig]{revtex}

\newcommand{\zaa}{Astron. Astrophys.}

\newcommand{\zapj}{Astrophys. J.}
\newcommand{\znp}{Nucl.~Phys.}

\newcommand{\zpr}{Phys.~Rev.}

\newcommand{\zppnp}{Progr.~Part.~Nucl.~Phys.}

\newcommand{\gap}{\mathrel{ \rlap{\raise.5ex\hbox{$>$}}
                    {\lower.5ex\hbox{$\sim$}}  } }
\newcommand{\lap}{\mathrel{ \rlap{\raise.5ex\hbox{$<$}}
	            {\lower.5ex\hbox{$\sim$}}  } }

\newcommand{\na}{$^{22}$Na}
\newcommand{\neo}{$^{22}$Ne}
\newcommand{\mg}{$^{22}$Mg}
\newcommand{\napg}{$^{21}$Na(p,$\gamma)^{22}$Mg}

\begin{document}
\twocolumn

\title{On the \napg\ thermonuclear rate for \na\ production in novae}

\author{Nadya A. Smirnova$^*$ and Alain Coc}

\address{
Centre de Spectrom\'etrie Nucl\'eaire et de Spectrom\'etrie
de Masse, IN2P3-CNRS and Universit\'e Paris Sud, B\^atiment 104,\\ 
91405 Orsay Campus, France
}

\maketitle

\begin{abstract}
Classical novae are potential sources of $\gamma$-rays, like the
1.275~MeV gamma emission following \na\ beta decay, that could be detected by 
appropriate instruments on board of future satellites like INTEGRAL.
It has been shown that the production of \na\ by novae is affected
by the uncertainty on the \napg\ rate and in particular by the unknown
partial widths of the $E_x$ = 5.714, $J^\pi$ = 2$^+$, \mg\ level. 
To reduce these uncertainties, we performed shell model calculations with
the OXBASH code, compared the results with available spectroscopic data and
calculated the missing partial widths.
Finally, we discuss the influence of these results on the \napg\ reaction 
rate and \na\ synthesis.
\end{abstract}



\section{Introduction}

Classical novae are potential sources of \na\ which beta decays
($\tau$ = 3.75~years) towards the first excited state of \neo\ followed
by a prompt gamma ray emission ($E_\gamma$=1.275~MeV) that could be
observed\cite{Cla74}.
Even though it has not been detected yet\cite{Iyu95}, this could happen with
higher sensitivity instruments on board of future space missions
like INTEGRAL (including a Ge based gamma--ray telescope\cite{SPI}) or
gamma--ray focusing telescope projects\cite{lens}.
If conditions are favorable (i.e. an oxygen--neon nova at a distance of less
than $\approx$2~kpc\cite{na}) the \na\ line could be detected\cite{Gom98} 
with the INTEGRAL spectrometer.
Nova outbursts occur at the surface of an accreting white dwarf within a
binary system. The accreted H--rich matter enriched with the C--O or
O--Ne matter from the white dwarf undergo a thermonuclear runaway that
synthesizes new isotopes. The formation of \na\ (in O--Ne novae) proceeds
from initial $^{20}$Ne present in large quantities through the two possible
paths : $^{20}$Ne(p,$\gamma)^{21}$Na($\beta^+)^{21}$Ne(p,$\gamma)^{22}$Na and
$^{20}$Ne(p,$\gamma)^{21}$Na(p,$\gamma)^{22}$Mg($\beta^+)^{22}$Na.
The first path has been found\cite{JCH99} to be more favorable to \na\
formation because of its longer time scale.
The preferred path is governed by the competition between the $^{21}$Na
$\beta^+$--decay and the \napg\ reaction whose rate remains uncertain mainly
because of the unknown resonance strength associated with the 
$E_x$ = 5.714, $J^\pi$ = 2$^+$, \mg\ level\cite{JCH99}.

Estimates of the $^{21}$Na(p,$\gamma)^{22}$Mg reaction rate\cite{CF88} have
been provided\cite{WGTR86,WL86}, considering the first three levels\cite{End90}
($E_x;\;J^\pi$ = 5.714; 2$^+$,  5.837; (0--5) and 5.965~MeV; 0$^+$)
above the proton threshold (5.501~MeV).
With the exception of the first level, the total widths can be identified
with the proton widths, so that $\omega\gamma\approx\omega\Gamma_\gamma$.
The resonance strength associated with the first $E_x$=5.714~MeV, 
$J^\pi$=2$^+$ level suffers from a significant uncertainty that affects the
\napg\ rate in the temperature domain of nova nucleosynthesis\cite{JCH99}.
For this level, only the total width is known experimentally
($\Gamma$=16.5$\pm$4.4~meV\cite{End90}) and to calculate the corresponding
resonance strength, estimated proton widths\cite{WGTR86,WL86} have been used
together with the relation $\Gamma$ = $\Gamma_p$ + $\Gamma_\gamma$.
The two estimates ($l_p=0$ and $\theta^2_p=0.01$\cite{WGTR86} or
$l_p=2$ and $\theta^2_p=0.5$\cite{WL86}) lead to similar values
($\gamma$= 3.4 or 3.8~meV) very close to the maximum value ($\Gamma$/4),
obtained when $\Gamma_\gamma$ = $\Gamma_p$ = $\Gamma$/2.
However, based on the data available for the \neo\ $E_x$=6.120~MeV analog
level\cite{End98}, it was argued\cite{JCH99} that the proton width could be
much smaller because {\it i)} the radiative width estimated from the analog
level is such that $\Gamma_\gamma\approx\Gamma$ and {\it ii)} the neutron
spectroscopic factor in the analog level should be very small according to
experimental data\cite{Neo72}.
Accordingly, values of $\omega\gamma$ = 2.5, 0.25, 0.0~meV have been adopted 
for upper ($\Gamma_\gamma$ = $\Gamma_p$ = $\Gamma$/2), recommended (with the
usual 0.1 reduction factor) and lower limit for the calculation of \na\
production in novae\cite{JCH99}. 
This induces a factor of 10$^5$ uncertainty on the rate around a temperature
of $\approx10^8$~K, typical of novae, and a factor of up to 3 in the \na\
yields\cite{JCH99}. Another, much less important, source of 
uncertainty\cite{JCH99} comes from the assumed value\cite{WGTR86,WL86} for 
the radiative width of the third ($E_x;\;J^\pi$ = 5.965~MeV; 0$^+$) level.    
It is important to reduce this uncertainty on the \na\ yield that directly 
affects the detectability distance of the 1.275~MeV gamma emission in order
to interpret future nova observations.
In consequence, we performed shell model calculations of spectroscopic
factors and radiative strengths for \neo\ and \mg\ nuclei. 
In this paper, we first compare calculated values with existing experimental
spectroscopic data in order to validate the calculations. 
We re--analyze existing experimental data to extract supplementary 
information on missing spectroscopic factors. 
From this analysis, we derive better estimates for the spectroscopic 
factor of the 5.714; 2$^+$ level and the radiative width of
the 5.965~MeV; 0$^+$ level. Finally, we discuss the influence
of these new values on the \napg\ rate and \na\ production in novae.

\section{Shell Model calculations}
\label{s:sm}

\begin{figure}
\epsfig{file=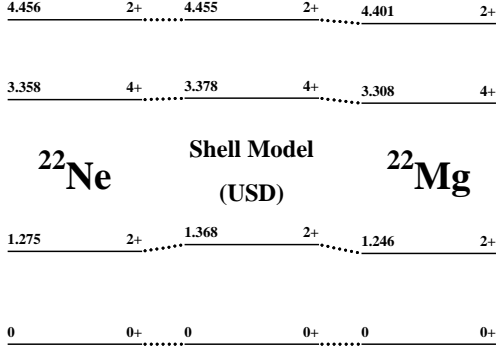,width=8.cm}
\caption{Positive parity energy levels of T=1 A=22 nuclei calculated with 
the USD interaction in comparison with the experimental level schemes of 
$^{22}$Ne and $^{22}$Mg up to 5~ MeV.}
\label{f:levelz}
\end{figure}

\begin{figure}
\epsfig{file=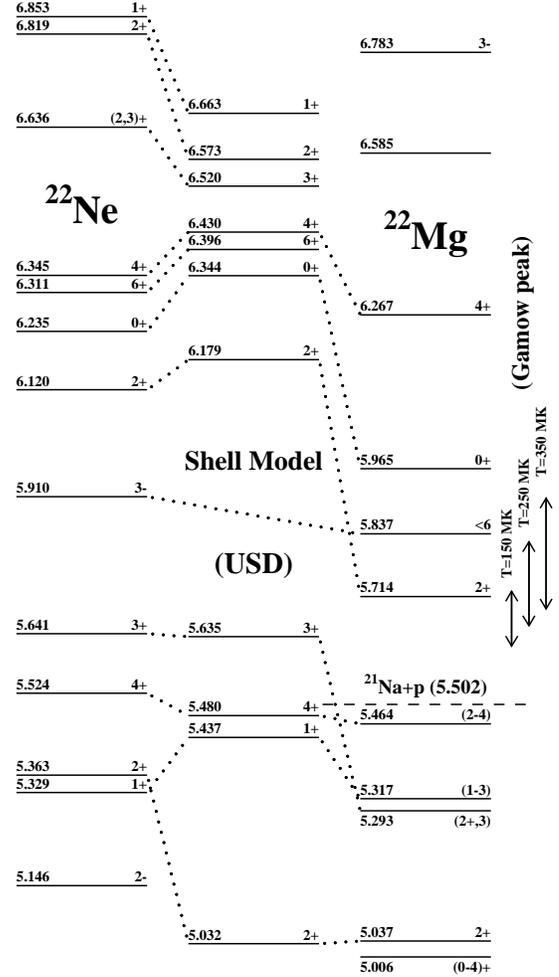,width=8.cm}
\caption{Same as Fig.~\protect\ref{f:levelz} but in the
energy range between 5 and  6.8~MeV.
In addition, for $^{22}$Mg, the threshold for proton emission is shown
(dashed line) together with the position of the Gamow peak for 
temperatures relevant to nova outbursts (arrows).
}
\label{f:levels}
\end{figure}

In order to estimate spectroscopic factors and radiative widths,
we have performed shell model calculations using the OXBASH code~\cite{OXBASH}.
Since we are interested only in positive parity states, we used the 
well-known USD interaction of Wildenthal~\cite{USD} for the $sd$ shell model 
space.
The results of the calculations in comparison with the experimental spectra
of $^{22}$Ne and $^{22}$Mg are shown in Figs.~\ref{f:levelz}, 
\ref{f:levels} and Table~\ref{t:spect}.
The correspondence of the experimental and theoretical results is remarkably 
good.

In Table~\ref{t:spect} the calculated neutron spectroscopic factors in 
\neo\ are compared with the experimental values obtained\cite{Neo72}
through the neutron stripping reaction $^{21}$Ne(d,p)$^{22}$Ne.
The agreement between calculated and experimental values is very
good except for the levels labeled (in ref.\cite{Neo72}) 6.120, 6.350; 6$^+$ 
(no experimentally determined spectroscopic factors)
and 7.341; 0$^+$.
However, these discrepancies can be explained when new spectroscopic
data\cite{End98} are included (see the following section).

Assuming the equality between spectroscopic factor of conjugate reactions
and according to the good agreement between our calculations and experiment
we can confidently use them to obtain the \mg\ proton width of the
$E_x;\;J^\pi$ = 5.714; 2$^+$ level.
The fourth calculated $2^+$ state at 6.179~MeV corresponds to the fourth $2^+$
state of $^{22}$Ne at 6.120~MeV and to the $2^+$ state of $^{22}$Mg at
5.714~MeV (Fig.~\ref{f:levels}) which are of main interest here.
To check the correctness of the assignment and the proximity of the calculated
and physical $2^+$ state, we compared calculated radiative strength with those
available experimentally\cite{ToI} for the $E_x$=6.120~MeV level in \neo.
There is a fair agreement between the experimental and theoretical values 
as shown in Table~\ref{t:gamma} increasing the confidence in the assignment.
The calculated spectroscopic factors for the 2$^+$ state are small as expected
from experiments on \neo\ (see Table~\ref{t:spect}). 
The corresponding proton reduced widths are obtained, using the relation
$\Gamma_{p} = C^2 \; S \; \Gamma_{s.p.}$ where the single-particle 
width $\Gamma_{s.p.}$ has been estimated from the scattering phase shifts
in the Woods-Saxon potential with the depth required to reproduce the
experimentally known energy of the resonance.
The contribution of $l=2$ transfer to the 5.714 MeV state in $^{22}$Mg
is negligible as compared to $l=0$ transfer and we obtain the values of
$\Gamma_{0d_{5/2}} = 2 \times 10^{-6}$ eV,
$\Gamma_{0d_{3/2}} = 2 \times 10^{-7}$ eV,
$\Gamma_{0s_{1/2}} = 4.5 \times 10^{-3}$ eV,
that leads to $\Gamma_p = 4.5$ meV.

We also obtained the radiative width of the third level above threshold
($E_x;\;J^\pi$ = 5.965~MeV; 0$^+$) by calculating $B(E2)$ or $B(M1)$ for the
transition to the lower lying $2^+$ and $1^+$ levels (Table~\ref{t:gamma0}).
The calculated value, $\Gamma_\gamma$ = 33.6~meV leads to a resonance
strength of $\omega\gamma$ = 4.2~meV, only slightly higher than the estimated
value \cite{WGTR86} of 2.5~meV.

\section{Reanalysis of experimental data}
\label{s:dwba}

We performed a DWBA analysis of some of Neogy et al.\cite{Neo72} data (6.120,
6.350 and 7.341 levels) to extend the comparison between calculated
and experimental values and to put more constraint on the 6.120~MeV \neo\ 
(or 5.714~MeV \mg) level spectroscopic factor. For this purpose, we used
the ECIS code\cite{ECIS} with the same optical potential parameters as 
Neogy et al.

The 6.350~MeV level (Fig.~5 of Neogy et al.\cite{Neo72}) was assumed to have
a 6$^+$ spin and parity and accordingly ($l=4$) no DWBA analysis was
performed at that time.
However, since then\cite{End98}, a 6$^+$ level has been located at 6.311~MeV
and a 4$^+$ level at 6.345~MeV. Hence, it is almost certain that the
6.350~MeV level in ref.\cite{Neo72} was unduly identified with the 6$^+$
instead of the 4$^+$ level. Accordingly, we made a DWBA analysis of the
Neogy et al. data assuming that the reported level is a 4$^+$ and extracted
a spectroscopic factor. This new value is close (within a factor of two) to
the calculated, 4$^+_3$ level, value (see Table~\ref{t:spect}). 
Hence, shell model calculations reinforce the idea that the 6.35~MeV level 
seen by Neogy et al.\cite{Neo72} is the 6.345; 4$^+$ one instead of the 
6.311; 6$^+$.

Our shell model calculations lead to a  very small spectroscopic factor for
the 7.341; $0^+$ level in complete contradiction with the value reported
by Neogy et al.
However, less than three keV above ($E_x$ = 7.344~MeV) lies a $J^\pi$ =
$(3,4)^+$ level whose calculated spectroscopic factors (for the 3$^+_3$ 
and 4$^+_4$ states) agree much better with those extracted from experimental
data (Table~\ref{t:spect}). 
Hence, the experimentally determined spectroscopic factor
could be attributed to the 7.344~MeV; $(3,4)^+$ level rather than to 
the 7.341; $0^+$ one.

\begin{figure}
\epsfig{file=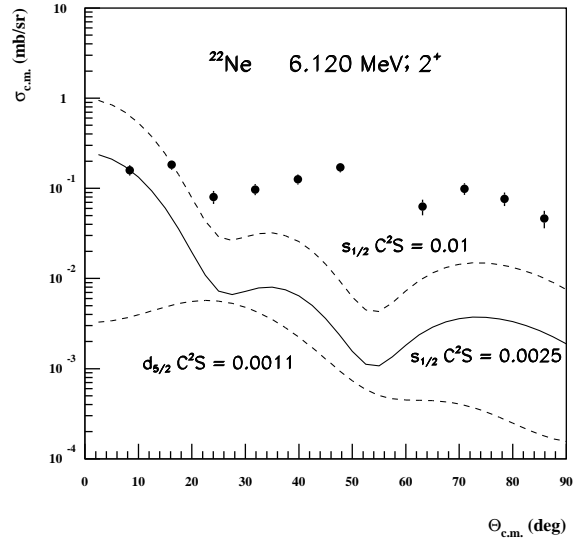,width=8.cm}
\caption{Angular distribution for the $^{21}$Ne(d,p)$^{22}$Ne neutron
transfer reaction on the 6.120~MeV; 2$^+$ level. The experimental data is 
from Neogy et al. \protect\cite{Neo72}. The curves represent DWBA
calculations with calculated spectroscopic factors (dashed lines) or
adjusted spectroscopic factors (solid line).}
\label{f:neogy}
\end{figure}

In order to put more constraint on the 6.120~MeV $^{22}$Ne level we also
performed a DWBA analysis of the Neogy et al. data for this level.
(This analysis was not performed in the original work\cite{Neo72} because
``the angular distribution does not exhibit characteristics of direct
reactions''.)
The experimental data and DWBA cross sections are represented in
Fig.~\ref{f:neogy}.
As expected, the
calculated transfer cross sections are more forward peaked than the
experimental angular distribution suggesting a strong contribution from
fusion reactions. When the theoretical spectroscopic factors are used, the
$d_{5/2}$ contribution is negligible while the $s_{1/2}$ contribution is
compatible with experimental data except for the most forward angle.
Requesting that the DWBA cross sections remain below all experimental data
points lead to upper limits for the spectroscopic factors of
$C^2S\;\lap\;0.0025$ or 0.015 for $l$ = 0 or 2 respectively.
While the $l$ = 2 upper limit is fully compatible with shell model
calculations, the $l$ = 0 one is a factor of four below the calculated
spectroscopic factor.
From this upper limit, we obtain $\Gamma_p$ $\lap$ 1~meV.
One can note however that this limit should be taken with caution as it
comes from a single data point at the smallest angle. As it can be seen in
Fig.~1 of Neogy et al., the 6.120~MeV, \neo\ peak is close to an other one
from $^{23}$Ne (arising from a (d,p) reaction on \neo\ in the target.)  
Resolving these two peaks at a lower angle should be more difficult
because of the unfavorable evolutions of both their energy separation 
and relative heights. Hence one cannot exclude that the experimental 
error bars were underestimated in this case. 

\section{Implications on the \napg\ rate}
\label{s:rate}

The contribution of the $E_x;\;J^\pi$ = 5.714; 2$^+$ level to the \napg\
rate depends directly from its adopted proton width. From experimental
data\cite{Neo72} we deduced an approximate upper limit (see comment
above) for this width ($\Gamma_p$ $\lap$ 1~meV). 
Our shell model calculations give a slightly higher value 
($\Gamma_p$ = 4.5~meV).
Using the experimental total width ($\Gamma$ = 16.5~meV), one obtains
$\gamma$ $\approx$ 3.2 or 1~meV, and $\omega\gamma$ = 2. or 0.6~meV, 
when using the calculated or experimental upper limit for the proton widths.
These values are close to the first (3.4\cite{WGTR86} and 3.8~meV\cite{WL86})
and more recent estimated strengths (2.5 upper and 0.25~meV nominal 
values\cite{JCH99}). 
One important conclusion resulting from the shell model calculations is that
it is most unlikely that the spectroscopic factor is much smaller than
one meV.
This would exclude the lower limit for the rate\cite{JCH99} obtained with a
null spectroscopic factor.

The contribution of the 5.965~MeV; 0$^+$ level is obtained through our 
shell model calculation of its radiative width
($\Gamma_\gamma$ = 33.6~meV and $\omega\gamma$ = 4.2~meV).
As in previous works\cite{WGTR86,WL86,JCH99} we assume that the 5.837~MeV;
(0-5) level, is the analog of the 5.910; 3$^-$ level in \neo.
This assignment made by \cite{End90}, is not present anymore
in\cite{End98} but is likely from the examination of Fig.~\ref{f:levels}
and because of their similar gamma decay modes\cite{ToI}. Accordingly, we
take $\Gamma_\gamma$ = 13~meV from the analog \neo\ level and $\omega\gamma$
= 11.4~meV. The direct capture term\cite{WGTR86,CF88} is also left unchanged
as it is based on experimental spectroscopic factors.

Following the good general agreement between calculated and experimental 
quantities, we derive the \napg\ rate using the shell model calculated 
values presented above.
The resonant part of the rate is approximated as usual by 
${\sum_i}A_i\exp(-B_i/T_9)$ with $A_i$ = 334., 1862., 686. for 
$B_i$ = 2.52, 3.95, 5.49 respectively. 
The resulting rate is very close to the previous rates \cite{WGTR86,WL86,CF88}
and of the recent upper rate limit\cite{JCH99} but it is now put on a safer 
ground, in the domain of nova nucleosynthesis, as it now relies on shell model 
calculations rather than estimates. 
Using the experimental upper limit instead of the calculated one for the
first resonance strength would lead to a rate lying within
a factor of three of the rates from refs.~\cite{WGTR86,WL86,CF88} and from 
the nominal and upper rate limit from ref.~\cite{JCH99}.
(The proximity of these different rates is due to the strong experimental 
constraint introduced by $\Gamma$ = $\Gamma_p$ + $\Gamma_\gamma$ =
16.5$\pm$4.4~meV\cite{End90}.)

In conlusion, the lower limit for the \napg\ rate, used for 
calculating\cite{JCH99} \na\ 
yields in novae seems now excluded. Unfortunately it was also the
more efficient for \na\ production through the 
$^{21}$Na($\beta^+)^{21}$Ne(p,$\gamma)^{22}$Na chain.
Hence the highest \na\ yields reported\cite{JCH99} are now excluded.
A precise conclusion on gamma emission detectability will require further
hydrodynamical calculations of nova outbursts.
However, the \napg\ rate resulting from this analysis is not too far from 
the nominal rate used in previous calculations\cite{JCH99} so that the nominal 
detectability distance of \na\ gamma emission\cite{na} should not be too 
much affected. 

\section{Acknowledgments}
It is a pleasure to thank J.-P.~Thibaud for constructive comments and 
J.~Kiener for usefull discussions and help in using the ECIS code.

\onecolumn

\mediumtext
\begin{table*}
\caption[]{
Experimental, positive parity,  energy levels of $^{22}$Mg and $^{22}$Ne and
spectroscopic factors deduced from the $^{21}$Ne(d,p)$^{22}$Ne reaction
compared with calculated ($T=1$ $A=22$) values.}
\begin{tabular}{cc|ccc|c|ccc}
\multicolumn{2}{c|}{$^{22}$Mg~$^a$}&\multicolumn{3}{c|}{$^{22}$Ne~$^b$}&&
\multicolumn{3}{c}{Shell Model}\\
\hline
$J^{\pi}$ & $E_x$ (MeV) &
$J^{\pi}$ & $E_x$ (MeV) & $(2J+1)S$~$^c$ & $l$ & 
$J^{\pi}$ &$E_x$ (MeV) & $(2J+1)S$ \\
\hline
$0^+$    & 0.    & $0^+$   & 0.    & $\le$ 0.20 & 2 &0$^+_1$& 0. & 0.13 \\
$2^+$    & 1.246 & $2^+$   & 1.275 & 3.25 & 2 &2$^+_1$& 1.368 & 4.9 \\
$4^+$($2^+$)&3.308 & $4^+$ & 3.358 & 0.44 & 2 &4$^+_1$& 3.378 & 0.29 \\
$2^+$($1^+$)&4.401 & $2^+$ & 4.456 & 0.27 & 0 &2$^+_2$& 4.455 & 0.30 \\
         &       &         &       & 0.72 & 2 &&      & 0.95 \\
$(0-4)^+$& 5.006 &         &       &      &   &       &       &      \\
(1,2)    & 5.317 & $1^+$   & 5.329 & 0.15 & 0 &1$^+_1$& 5.437 & 0.15  \\
         &       &         &       & 1.40 & 2 &       &      & 1.8 \\
$2^+$    & 5.037 & $2^+$   & 5.363 & 1.56 & 0 &2$^+_3$& 5.032 & 1.22 \\
(2,3,4)  & 5.464 & $4^+$   & 5.524 & 2.26 & 2 &4$^+_2$& 5.480 & 2.87 \\
($2^+$,3)& 5.293 & $3^+$   & 5.641 & 0.49 & 2 &3$^+_1$& 5.635 & 1.16 \\
(0-5)    & 5.837 &         &       &      &   &       &       &      \\
$2^+$    & 5.714 & $2^+$   & 6.120 &$\lap$0.012~$^d$& 0 &
2$^+_4$& 6.179 & 0.05 \\
         &       &         &       & $\lap$0.07~$^d$  & 2 &&       & 0.006\\
$0^+$    & 5.965 & $0^+$   & 6.235 &      & 2 &0$^+_2$& 6.344 & 0.05  \\
         &       & $6^+$   & 6.311 &      &   &6$^+_1$& 6.396 &     \\
$4^+$    & 6.267 & $4^+$   & 6.345 &$\approx$0.5~$^{d,e}$& 2 &
4$^+_3$& 6.430 & 1.0  \\
         &       &$(2,3)^+$& 6.636 & 0.72 & 2 &3$^+_2$& 6.520 & 0.92 \\
         &       & $2^+$   & 6.819 & 0.92 & 0 &2$^+_5$& 6.573 & 0.66 \\
         &       & $1^+$   & 6.854 & 1.65 & 0 &1$^+_2$& 6.663 & 1.65 \\
         &       & $0^+$   & 7.341 & (0.35)~$^f$ & 2 &0$^+_3$& 7.264 & 0.007 \\
         &       & $(3,4)^+$   & 7.344 & $\approx$0.48~$^{d,f}$
	 &2 &3$^+_3$& 7.742 & 1.1 \\
         &       & &&      $\approx$0.46~$^{d,f}$&2 &4$^+_4$& 6.993 & 0.28 \\
         &       &$(3,5)^+$ & 7.423 &      &   &5$^+_1$& 7.461 &  \\
         &       & $2^+$   & 7.644 & 0.12 & 0 &2$^+_6$& 7.804 & 0.7 \\
         &       &         &       & 0.41 & 2 &       &       & 1.4 \\
\end{tabular}
$^a$ $E_x$ and $J^\pi$ from Ref.~\cite{End90}\\
$^b$ $E_x$ and $J^\pi$ from Ref.~\cite{End98}\\
$^c$ $(2J+1)S$ from Ref.~\cite{Neo72} unless othewise stated.\\
$^d$ Our analysis of Neogy et al. data\protect\cite{Neo72}.\\
$^e$ Assuming that the 6.35; 6$^+$ \neo\ data in Neogy et al.
\protect\cite{Neo72} corresponds to the 6.345; 4$^+$ level.\\
$^f$ Assuming that the 7.341; 0$^+$ \neo\ data in Neogy et al.
\protect\cite{Neo72} corresponds to the 7.344; (3,4)$^+$ level.
\label{t:spect}
\end{table*}

\narrowtext
\begin{table}
\caption{Experimental and theoretical reduced transition
probabilites from 6.120 MeV $2^+$ state $^{22}$Ne.}
\begin{tabular}{ccc}
Transition&Experimental~$^a$&Theoretical\\
\hline
$2^+ \to 0^+_1$ & $B(E2)=(1.91\pm 1.11)$ e$^2$fm$^4$ &
$B(E2)=3.1$ e$^2$fm$^4$ \\
$2^+ \to 2^+_1$ & $B(E2)=(1.15\pm 1.15)$ e$^2$fm$^4$ & 
$B(E2)=1.8$ e$^2$fm$^4$ \\
$2^+ \to 2^+_1$ & $B(M1)=(0.046\pm 0.023)$ $\mu_N^2$& 
$B(M1)=0.138$ $\mu_N^2$ \\
\end{tabular}
$^a$ From Ref.~\cite{ToI}
\label{t:gamma}
\end{table}

\narrowtext
\begin{table}
\caption{Theoretical reduced transition probabilites
from $0^+_2$ state of which is assumed to correspond to the
5.965 MeV $0^+$ level in $^{22}$Mg.}
\begin{tabular}{cc}
Transition &Transition rates \\
\hline
$0^+_2 \to 2^+_1$ & $B(E2)=16.92 $ e$^2$fm$^4$ \\
$0^+_2 \to 2^+_2$ & $B(E2)=8.0$ e$^2$fm$^4$ \\
$0^+_2 \to 2^+_3$ & $B(E2)=6.2$ e$^2$fm$^4$ \\
$0^+_2 \to 2^+_4$ & $B(E2)=0.01$ e$^2$fm$^4$ \\
$0^+_2 \to 1^+_1$ & $B(M1)=0.57$ $\mu_N^2$ \\
\end{tabular}
\label{t:gamma0}
\end{table}

\end{document}